\documentclass[aps,floatfix,prd,nofootinbib,superscriptaddress,reprint,showpacs,10pt,preprintnumbers,longbibliography,showdoi]{revtex4-1}
\usepackage[utf8]{inputenc}
\usepackage[pdftex]{graphicx}
\usepackage{float}
\usepackage{amsmath}
\usepackage{amssymb}
\usepackage{mathtools}
\usepackage{siunitx}
\usepackage{amsfonts}
\usepackage{dsfont}
\usepackage{array}
\usepackage{bm}
\usepackage{mathrsfs}
\usepackage{pifont}
\usepackage{multirow}
\usepackage{upgreek}
\usepackage[dvipsnames]{xcolor}
\usepackage[pdftex,
  pdftitle={Digitising SU(2) Gauge Fields and the Freezing Transition},
  pdfauthor={T. Hartung, T. Jakobs, K. Jansen, J. Ostmeyer, C. Urbach},
  bookmarks,
  colorlinks,
  linkcolor=myblue,
  citecolor=mymagenta,
  menucolor=black,
  urlcolor=myblue,
  plainpages=false,
  pdfpagelabels,
  hypertexnames=false]{hyperref}
\usepackage{verbatim}
\usepackage{slashed}
\usepackage{cleveref}
\usepackage{ucs}
\usepackage{subfigure}

\usepackage{tikz}

\newcommand{\ri}{\mathrm{i}}
\newcommand{\SUTwo}{SU$(2)$}
\newcommand{\abs}[1]{|#1|}
\newcommand{\matr}[2]{\left(\begin{array}{#1}#2\end{array}\right)}

\DeclareMathOperator{\ord}{\mathcal{O}}

\newcommand{\ordnung}[1]{\ensuremath{\ord\left(#1\right)}}

\definecolor{mymagenta}{RGB}{200, 0, 100}
\definecolor{myblue}{RGB}{45, 48, 146}

\graphicspath{{plots/}}

\begin{document}
\title{Digitising SU$(2)$ Gauge Fields and the Freezing Transition}

\author{Tobias Hartung}
\affiliation{Department of Mathematical Sciences, University of Bath, 4 West, Claverton Down, Bath, BA2 7AY, UK}
\affiliation{Computation-based  Science  and  Technology  Research  Center, The  Cyprus  Institute,  20  Kavafi  Street,  2121  Nicosia, Cyprus}

\author{Timo Jakobs}
\affiliation{Helmholtz-Institut f\"ur Strahlen- und Kernphysik, University of Bonn, Nussallee 14-16, 53115 Bonn, Germany}
\affiliation{Bethe Center for Theoretical Physics, University of Bonn, Nussallee 12, 53115 Bonn, Germany}

\author{Karl Jansen}
\affiliation{NIC, DESY Zeuthen, Platanenallee 6, 15738 Zeuthen, Germany}

\author{Johann Ostmeyer}
\affiliation{Department of Mathematical Sciences, University of Liverpool, Liverpool, L69 7ZL, United Kingdom}

\author{Carsten Urbach}
\affiliation{Helmholtz-Institut f\"ur Strahlen- und Kernphysik, University of Bonn, Nussallee 14-16, 53115 Bonn, Germany}
\affiliation{Bethe Center for Theoretical Physics, University of Bonn, Nussallee 12, 53115 Bonn, Germany}

\date{\today}

\begin{abstract}
  Efficient discretisations of gauge groups are crucial with the long
  term perspective of using tensor networks or quantum computers for lattice gauge theory
  simulations. For any Lie group other than U$(1)$, however, there is
  no class of asymptotically dense discrete subgroups. Therefore,
  discretisations limited to subgroups are bound to lead to a
  \textit{freezing} of Monte Carlo simulations at weak couplings,
  necessitating alternative partitionings without a group structure. In
  this work we provide a comprehensive analysis of this freezing for
  all discrete subgroups of SU$(2)$ and different classes of
  asymptotically dense subsets. We find that an appropriate choice of
  the subset allows unfrozen simulations for arbitrary couplings,
  though one has to be careful with varying weights of unevenly
  distributed points. A generalised version of the Fibonacci spiral
  appears to be particularly efficient and close to optimal.
\end{abstract}

\maketitle

\section{Introduction}

Gauge theories represent the main ingredients to the current standard
model (SM) of particle physics, which unifies the electromagnetic, the
weak and the strong interactions. Despite the tremendous success of
the SM, first principle calculations in non-Abelian gauge theories
underlying for instance quantum chromodynamics (QCD), which describes
the strongly interacting part of the SM, are still challenging. In the
last decades lattice field theoretical methods have been developed and
optimised with great success to provide a non-perturbative approach
for the investigation of such gauge theories using Monte Carlo (MC)
methods.

However, studying QCD for instance at finite density or its real time 
dynamics is difficult if not impossible with MC methods, either due to
the sign problem or because Euclidean space-time is
used. Here is where methods based on the Hamiltonian formalism in
Minkowski space-time can provide a way out. In fact, tensor network
methods have seen very rapid developments in the recent years towards
the possibility of simulations in $2+1$ and $3+1$ space-time
dimensions \cite{Schneider2021,Montangero2021}. And the number of qubits available on real quantum devices
is ever increasing. This offers a prospect for studying gauge theories
with tensor network methods or on quantum computers in the not too
distant future.

The Hamiltonian formalism for non-Abelian gauge theories with or
without matter content was presented a long time ago in
Ref.~\cite{Kogut:1974ag}. Its implementation with TN methods or on a
quantum computer, however, requires some form of digitisation of
SU$(N)$.

There are different ways to digitise SU$(N)$ or more specifically
SU$(2)$ which we will study in this paper. One can for instance chose
a discrete  subgroup of SU$(2)$. In the early days of lattice gauge
theory simulations such discrete subgroups were already investigated to
improve the efficiency of the simulation programmes. Soon it was
realised that due to the finite number of elements in such subgroups a
so-called freezing phase transition occurs at some critical
$\beta$-value~\cite{Petcher:1980cq,Jansen:1985nh}. For $\beta$-values larger than
this critical value MC simulations are no longer reliable, because
they result in the wrong distribution. For SU$(3)$ a particular choice
of digitisation was studied in
Refs.~\cite{Alexandru:2019nsa,Ji:2020kjk,Alexandru:2021jpm}.

While discrete subgroups have the advantage that they are closed under
multiplication, there is no flexibility in the number of group
elements. This motivates using the isomorphy between SU$(2)$ and the
sphere $S_3$ in four dimensions. The aim is then to find points on
$S_3$ depending on some parameter $m$ which are dense in $S_3$ as $m$
approaches infinity. 

In this paper we investigate all the discrete subgroups of
SU$(2)$ and several representative discretisations of $S_3$. We study
the freezing transition as a function of the number of elements in
these discretisations and show that the discretisation based on
so-called Fibonacci lattices behaves optimally.

\section{Lattice Action}

We work on a hypercubic, Euclidean lattice with the set of lattice sites
\[
\Lambda\ =\ \{n=(n_0,\ldots, n_{d-1})\in\mathbb{N}_0^d: n_\mu = 0, 1,
\ldots, L-1\}\,, 
\]
with $L\in\mathbb{N}$. At every site there are $d\geq 2$ link variables
$U_\mu(n)\in\mathrm{SU}(2)$ connecting to sites in forward direction
$\mu=0, \ldots , d-1$. We define the plaquette operator as
\begin{equation}
  P_{\mu\nu}(n)\ =\ U^{~}_\mu(n) U^{~}_\nu(n+\hat\mu)
  U^\dagger_\mu(n+\hat\nu) U^\dagger_\nu(n)\,,
\end{equation}
where $\hat\mu\in\mathbb{N}_0^d$ is the unit vector in direction
$\mu$. In terms of $P_{\mu\nu}$ we can define Wilson's lattice
action~\cite{Wilson:1974sk} 
\begin{equation}
S = -\frac{\beta}{2}\sum_n \sum_{\mu<\nu} \mathrm{Re}\,\mathrm{Tr}\,P_{\mu \nu}(n)\,,
\end{equation}
with $\beta$ the inverse squared gauge coupling. We will use the
Metropolis Markov Chain Monte Carlo algorithm to generate chains of
sets $\mathcal{U}_i$ of link variables $\mathcal{U} = \{U_\mu(n):
n\in\Lambda, \mu=0,\ldots, d-1\}$ distributed according to
\begin{equation}
  \mathbb{P}(\mathcal{U})\ \propto\ \exp[-S(\mathcal{U})]\,.
\end{equation}
The main observable we will study in this paper is the plaquette 
expectation value defined as
\begin{equation}
  \langle P\rangle = \frac{1}{N}\sum_{i=1}^N\ P(\mathcal{U}_i)
\end{equation}
with
\[
P(\mathcal{U}) = \frac{2}{d(d-1) L^d}\sum_n \sum_{\mu<\nu}
\mathrm{Re}\,\mathrm{Tr}\,P_{\mu \nu}(n)\,. 
\]

\section{SU$(2)$ Partitionings}

In Monte Carlo simulations of lattice SU$(N)$ gauge theories using the
Metropolis algorithms or some variant of ti one typically
requires a proposal gauge link at site $n$ in direction $\mu$ obtained
as
\[
U'_\mu(n) = V\cdot U_\mu(n)\,.
\]
Here, $V$ is a random element of SU$(N)$ with average distance
$\delta$ to the identity element. The average distance, measured using
some norm, determines the acceptance rate of the MC algorithm.

The actual value of $\delta$ needs to be adjusted to tune the
acceptance rate to about 50\%, which implies that for $\beta\to\infty$
one needs to decrease $\delta$ further and further.

In numerical simulations, one nowadays represents an element $U$ of
SU$(N)$ by an $N\times N$ complex valued  matrix and constrains it to
be unitary with unit determinant. Every complex number is then
represented by two floating point numbers with accuracy limited by the
adopted data type (usually double precisions floating point
numbers). The results obtained with this quasi continuous
representation of SU$(2)$ will be referred to as \emph{reference} results in
the following.
It imposes, for $\beta$-values of practical relevance, no
restriction on the possible elements $U'$: small enough distances
$\delta$ are possible.

However, this is not necessarily the case if a finite set of elements
of SU$(N)$ is to be used, like for instance a finite subgroup of
SU$(N)$. Here, there is a lower bound for the distance between two
available elements, which significantly restricts the possible
proposal gauge links. For too large $\beta$-values, therefore, the
acceptance drops to (almost) zero, an effect that was dubbed freezing
transition~\cite{Petcher:1980cq}.

This transition can be pushed towards larger and larger $\beta$-values
by increasing the number of elements in the set. Since there are in
general no finite subgroups of SU$(N)$ with arbitrarily many elements
available, one needs to resort to sets of elements which do not form a
subgroup of SU$(N)$, but which lie asymptotically dense and are
as isotropically as possible distributed in the group. We will call
these sets \emph{partitionings} of SU$(N)$.

Focusing on SU$(2)$, we discuss first some finite subgroups followed
by other partitionings of SU$(2)$.

\subsection{Finite Subgroups of \SUTwo}

Due to the double cover relation between SU$(2)$ and SO$(3)$, finite
subgroups of SU$(2)$ can be
constructed by taking the Cartesian product of the cyclic group of order 2 with
the subgroups of SO$(3)$. Subgroups of SO$(3)$ are obtained by considering the
symmetry transformations of regular polygons, as well as the rotational
symmetries of platonic solids~\cite{Klein:1880}.

In the following we will consider the binary tetrahedral group
$\overline{T}$, the binary octahedral group $\overline{O}$ and the binary
icosahedral group $\overline{I}$, with $24$, $48$ and $120$ elements respectively.
Their elements are evenly distributed across the whole group, and research on their
behavior has already been conducted~\cite{Petcher:1980cq}.
The last four-dimensional finite subgroup of SU$(2)$ is the binary dihedral
group $\overline D_4$ with 8 elements.
One possible representation of these groups can be found in \cref{tab:subgroups}.

\begin{table*}[!hbt]
 \centering
 \begin{tabular}{c|c|l}
  group          & order & elements                                                                                                                                                   \\
  \hline
  $\overline{D}_4$ & $8$    & $\pm 1, \pm i, \pm j, \pm k$ \\
  \hline
  $\overline{T}$   & $24$    & all sign combinations of $\left\{\pm 1, \pm i, \pm j, \pm k, \frac{1}{2}\left( \pm 1 \pm i \pm j \pm k \right) \right\} $                                  \\
  \hline
  $\overline{O}$   & $48$    & all sign combinations and permutations of                                                                                                                  \\
                   &         & $\left\{\pm 1, \pm i, \pm j, \pm k, \frac{1}{2}\left( \pm 1 \pm i \pm j \pm k \right) , \frac{1}{\sqrt{2}}\left( \pm 1 \pm i \right) \right\} $            \\
  \hline
  $\overline{I}$   & $120$   & all sign combinations and even permutations of                                                                                                             \\
                   &       & $\left\{\pm 1, \pm i, \pm j, \pm k, \frac{1}{2}\left( \pm 1 \pm i \pm j \pm k \right) , \frac{1}{2}\left(1+\varphi  i + \frac{j}{\varphi} + 0k \right) \right\} $ \\
 \end{tabular}
 \caption{Quaternionic representation of $\overline{D}_4$, $\overline{T}$, $\overline{O}$ and $\overline{I}$ as found in \cite{duval:1964}, where $\varphi = \frac{1+\sqrt{5}}{2}$ denotes the golden ratio.}
 \label{tab:subgroups}
\end{table*}

\subsection{Asymptotically dense partitionings of SU$(2)$}\label{sec:dense_partitionings}

For generating partitionings of SU$(2)$, we use
the isomorphy between SU$(2)$ and the sphere $S_3$ in four
dimensions, which is defined by
\begin{equation}
  x\in S_3\ \Leftrightarrow\
  \begin{pmatrix}
    x_0 + \ri x_1 & x_2 + \ri x_3 \\
    -x_2 + \ri x_3 & x_0 - \ri x_1\\
  \end{pmatrix}\in \mathrm{SU}(2)\,.
\end{equation}
For such partitionings, the number of elements can
be increased very easily, i.e.\ the discretisation of SU$(2)$ can be
made arbitrarily fine.

The reduction approach of SU$(2)$ to a sphere can be generalised to general U$(N)$ and SU$(N)$ which can be expressed as products of spheres. To this end, we note that U$(1)$ is isomorphic to $S_1$ and $ \text{U}(N)\cong \text{SU}(N)\rtimes \text{U}(1)$ where $\rtimes$ denotes the semi-direct product. This follows from the existence of the short exact sequence $1\to\text{SU}(N)\to\text{U}(N)\stackrel{\det}{\longrightarrow}\text{U}(1)\to1$.

With respect to SU$(N)$,\footnote{For more detail, see e.g.\ chapter 22.2c in \cite{frankel}.} we note that SU$(N)$ acts transitively on $S_{2N-1}$ since the point $(1,0,0,\ldots,0)$ is mapped to a point $z$ by any element of SU$(N)$ whose first column is $z$, and the isotropy subgroup of $(1,0,0,\ldots,0)$ is the SU$(N-1)$ embedding
\begin{align*}
  \begin{pmatrix}
    1&0\\0&\text{SU}(N-1)
  \end{pmatrix}.
\end{align*}
Hence, we obtain
\begin{align*}
  \text{SU}(N-1)\to\text{SU}(N)\to\text{SU}(N)/\text{SU}(N-1)\cong S_{2N-1}
\end{align*}
which implies that SU$(N)$ is a principal bundle over $S_{2N-1}$ with fibre SU$(N-1)$. Thus, by induction with SU$(2)\cong S_3$, we can express SU$(N)$ as a product of odd-dimensional spheres $S_3$, $S_5$, $\ldots$, $S_{2N-1}$, and U$(N)$ as a product of odd-dimensional spheres $S_1$, $S_3$, $\ldots$, $S_{2N-1}$.

Our aim is therefore to find a discretisation scheme of the $k$-dimensional
sphere $S_k$ depending on some parameter $m$ so that the discretising
set $S^m_k$ is dense in $S_k$ as $m$ goes to infinity. The following
examples all meet this requirement. Yet they differ in the measure or
probabilistic weight attributed to each point. This measure $w$ is
defined as the volume of the Voronoi cell~\cite{Voronoi:1908a,Voronoi:1908b} of
the point using the canonical metric on $S_k$ derived from the
Euclidean distance, i.e.\ the measure is the volume of that part of
the sphere closer to the given point than to any other point.

\subsubsection{Genz points}

A first, quite intuitive, partitioning is given by the Genz points~\cite{genz_points} setting $S^m_k=G_m(k)$ where we define
\begin{align}
  \begin{split}
    G_m(k) &\coloneqq \left\{\left(s_0\sqrt{\frac{j_0}{m}},\dots,s_k\sqrt{\frac{j_k}{m}}\right)\right.\\
    \qquad&\left|\,\sum_{i=0}^kj_i=m,\;\forall i\in \{0,\dots,k\}:\,s_i\in\{\pm1\},\,j_i\in\mathbb{N}\right\}\,,
  \end{split}
\end{align}
that is all integer partitions $\{ j_0,\dots,j_k\}$ of $m\ge1$ including all permutations and adding all possible sign combinations. Whenever the argument is dropped, we implicitly set $k=3$. The nearest neighbours of a Genz point can be found (up to sign changes) by choosing all pairs $i,l\in\{0,\dots,n\}$ with $j_i>0$ and $j_l<n$ and replacing
\begin{align}
  j_i\mapsto j_i-1\,,\quad j_l\mapsto j_l+1\,.\label{eq:genz_nearest_neighbours}
\end{align}
Note that all elements of a Genz point other than the $i$-th and
$l$-th components remain unchanged by such a replacement because the
denominator $\sqrt{m}$ is fixed.
The square distance between neighbouring points reads
\begin{equation}
  \label{eq:sqdist}
  d(j_i,j_l)^2 = \left|\left(\sqrt{\frac{j_i}{m}}-\sqrt{\frac{j_i-1}{m}},\,\sqrt{\frac{j_l}{m}}-\sqrt{\frac{j_l+1}{m}}\right)\right|^2\,,
\end{equation}
which can be evaluated to
(see \Cref{sec:distances_derivation} for details)
\begin{equation}
  d(j_i,j_l)^2 = \frac 1m \left(\frac{1}{4j_i}+\frac{1}{4j_l}+\ordnung{j_i^{-2}}+\ordnung{j_l^{-2}}\right)\,,
\end{equation}
which is highly anisotropic. In the regions where both $j_i$ and $j_l$
are of the order $m$ the distance scales as $d\sim \frac 1m$ whereas
smaller values of $j_i$ and $j_l$ lead to $d\sim \frac{1}{\sqrt
  m}$. The minimal distance is $d(\frac m2,\frac m2)=\frac 1m
+\ordnung{m^{-3/2}}$ and the maximum is reached at
$d(1,0)=\sqrt\frac{2}{m}$. We thus find that the distance does not
only depend on the position of the point but also on the choice of the
neighbour. Therefore even an approximation of the measure $w$ would
require a product over the distances of a given point to all its
neighbours.

As a concluding remark we note that in $k$ dimensions the weight of
different point differs by a factor up to
\begin{align}
  \frac{w_\text{max}}{w_\text{min}} \sim m^{k/2}
\end{align}
where the least density (largest measure) is reached where many of the $j$s are zero. This is in particular the case near the poles.

\subsubsection{Linear discretisation}

In order to avoid the aforementioned anisotropy, we consider the
following, linearly discretised, set of points
\begin{align}
  \begin{split}
    L_m(k) &\coloneqq \left\{\frac{1}{M}\left(s_0j_0,\dots,s_kj_k\right)\right.\\
    &\left|\,\sum_{i=0}^kj_i=m,\;\forall i\in \{0,\dots,k\}:\,s_i\in\{\pm1\},\,j_i\in\mathbb{N}\right\}\,,
  \end{split}\label{eq:linear_discretisation}\\
  M &\coloneqq \sqrt{\sum_{i=0}^kj_i^2}\,.\label{eq:define_M}
\end{align}
$M$ takes values $m\geq M \geq \frac{m}{\sqrt{k+1}}$. The lower bound
is assumed when all the $j$ are equal and the upper one when all but
one $j$ are zero. Note that $L_1$ happens to coincide with the finite
subgroup $\bar{D}_4$.

We find the nearest neighbours as
before~\cref{eq:genz_nearest_neighbours} and we obtain the change in
$M$ from neighbour to neighbour as
\begin{equation}
  \Delta M = \frac{j_l-j_i}{M} + \ordnung{\frac 1m}
\end{equation}
yielding the inverse change
\begin{equation}
  \begin{split}
    \frac 1M - \frac{1}{M+\Delta M} &= \frac{\Delta M}{M^2}+\ordnung{\frac 1{M^3}}\\
    &= \frac{j_l-j_i}{M^3} + \ordnung{\frac 1{m^3}}\,.\\
  \end{split}
\end{equation}
With this we can again calculate the square distance (with an
equivalent definition to \cref{eq:sqdist}, for details see
\cref{sec:distances_derivation})
\begin{equation}
  d(j_i,j_l)^2 = \frac{(j_l-j_i)^2}{M^4}+\frac{2}{M^2}+\ordnung{\frac{1}{m^3}}\,.
\end{equation}
It follows from $|j_l-j_i|\le M$ that $\frac{\sqrt{2}}{M}\le d \le
\frac{\sqrt{3}}{M}$ to leading order. Thus the distance has only a
weak dependence on the direction and it always scales as $d\sim \frac
1m$ with a difference of at most a factor $\sqrt{k+1}$ between different
points. This difference is governed by the range of $M$. We therefore
find the largest density of points (smallest distance) with the
largest values of $M$ at the poles.

A good approximation for the weights is given by
\begin{align}
  \label{eq:linweights}
  w \approx \left(\frac{\sqrt{2}}{M}\right)^k
\end{align}
with the largest deviation
\begin{align}
  \frac{w_\text{max}}{w_\text{min}} \sim (k+1)^{k/2}\,.
\end{align}

\subsubsection{Volleyball}

A third partitioning reminds of a Volleyball. It is the class of geodesic polytopes~\cite{pugh1976polyhedra} simplest to construct with its points
given by
\begin{align}
  \begin{split}
    V_m(k) &\coloneqq \left\{\frac{1}{M} \left(s_0 j_0, \dots, s_k j_k \right) \right.\\
    & \left|\,  (j_0, \dots, j_k) \in \left\{ \text{all perm. of } (m, a_1, \dots, a_k)\right\}, \right. \\
    & \left. \,s_i\in\{\pm1\}, \, a_i \in \{0, \dots, m\} \vphantom{\frac 1M}\right\}
  \end{split}
\end{align}
with $M$ defined in \cref{eq:define_M}, which takes values $m \le M \le  \sqrt{k+1} \, m$.

Additionally, the corners of the hypercube, in four dimensions also called $C_8$, form
\begin{align}
	V_0(k) &\coloneqq \left\{\frac{1}{\sqrt{k+1}} \left(s_0, \dots, s_k \right) | \,s_i\in\{\pm1\}\right\}\,.
\end{align}

For $m\ge1$ nearest neighbours can be obtained by $j_i \pm 1$, as long as
the conditions from above hold. The corresponding change in $M$
is computed to
\begin{equation}
  \Delta M = \frac{\pm j_i}{M} + \ordnung{\frac 1m}\,,
\end{equation}
yielding the inverse change
\begin{align}
  \begin{split}
    \frac 1M - \frac{1}{M+\Delta M} &= \frac{\Delta M}{M^2}+\ordnung{\frac 1{M^3}}\\
    &= \frac{\pm j_i}{M^3} + \ordnung{\frac 1{m^3}}\,.\\
  \end{split}
\end{align}
The square distance in this case reads (see again
\cref{sec:distances_derivation} for details)
\begin{equation}
  d(j_i, j_l)^2 = \frac{j_i^2}{M^4}+\frac{1}{M^2}+\ordnung{\frac{1}{m^3}}\,,
\end{equation}
where from $|j_i|\le M$ follows that $\frac{1}{M}\le d \le
\frac{\sqrt{2}}{M}$ to leading order. Thus, like for the linear
partitioning $L_m(k)$, the distance has only a weak direction
dependence and it always scales as $d\sim \frac 1m$ with a difference
of at most a factor $\sqrt{k+1}$ between different points. This
difference is governed by the range of $M$. We therefore find the
largest density of points (smallest distance) with the largest values
of $M$ at the poles.
Then, a good approximation for the weights is given by
\begin{align}
  \label{eq:Vweights}
  w \approx \left(\frac{1}{M}\right)^k
\end{align}
with the largest deviation
\begin{align}
  \frac{w_\text{max}}{w_\text{min}} \sim (k+1)^{k/2}\,.
\end{align}

\subsubsection{Fibonacci Lattice}

\begin{figure*}
  \centering
  \includegraphics[width=0.8\textwidth]{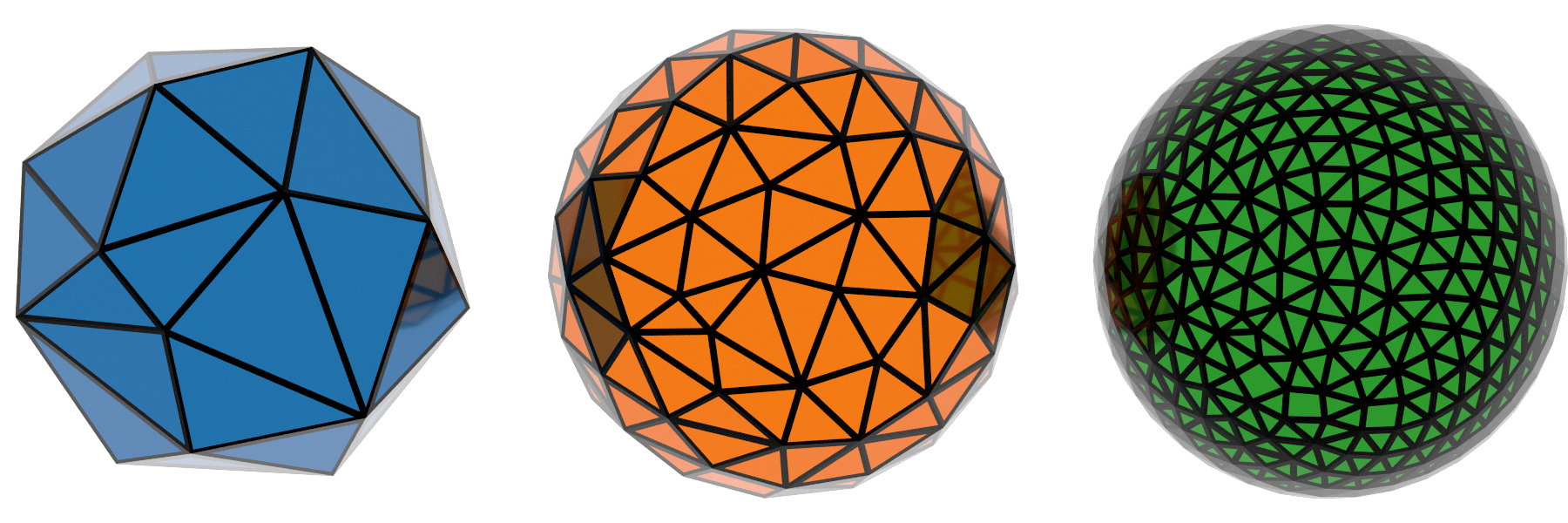}
  \caption{Fibonacci lattices on $S_2$ with $20$ (blue), $100$ (orange) and $500$ (green) vertices.}
  \label{fig:fibonacciPic}
\end{figure*}

The final discretization of \SUTwo\ considered in this work is a
higher dimensional version of the so-called Fibonacci lattice. It
offers an elegant and deterministic solution to the problem of
distributing a given amount of points on a two-dimensional
surface. Fibonacci lattices are used in numerous fields of research such as
numerical analysis or computer graphics, mostly to approximate spheres
(as e.g.\ shown in \cref{fig:fibonacciPic}). Mainly inspired by
\cite{stack:fibonacci}, we will now construct a similar
lattice for $S_3$.

The two-dimensional Fibonacci lattice is usually constructed within a
unit square $[0,1)^2$ as
\begin{align*}
  \Lambda_n^2                     & = \left\{ \tilde{t}_m \middle| 0 \le m < n, \, \, m \in \mathbb{N} \right\}                                    \\
  \textrm{with} \qquad \tilde{t}_m & = \begin{pmatrix}x_m\\y_m\end{pmatrix} = \left(\frac{m}{\tau} \quad \mathrm{mod} \quad 1, \frac{m}{n} \right)^t\,,\\
  \tau &= \frac{1+\sqrt{5}}{2}\,.
\end{align*}
This can be generalized to the hypercube $[0,1)^k$ embedded in $\mathbb{R}^k$:
\begin{align*}
 \Lambda_n^k & = \left\{ t_m \middle| 0 \le m < n, \, \, m \in \mathbb{N} \right\}                                                                                       \\
 t_m & = \begin{pmatrix} t_m^1 \\ t_m^2 \\ \vdots \\ t_m^k \end{pmatrix} = \begin{pmatrix}
  \frac{m}{n}        &                      \\
  a_1 \, m \quad     & \mathrm{mod} \quad 1 \\
  \vdots             &                      \\
  a_{k-1} \, m \quad & \mathrm{mod} \quad 1 \\
 \end{pmatrix}
\end{align*}
with
\[
\frac{a_i}{a_j} \notin \mathbb{Q} \quad \textrm{for} \quad i \neq j \textrm{,}
\]
where $\mathbb{Q}$ denotes the field of rational numbers. The square roots of the prime numbers provide a simple choice for the constants $a_i$:
\begin{align*}
 (a_1, a_2 ,a_3, \dots) = (\sqrt{2}, \sqrt{3}, \sqrt{5}, \dots)
\end{align*}
The points in $\Lambda_n^k$ are then evenly distributed within the given Volume. All that is left to do is to map these points onto a given compact manifold $M$, in our case \SUTwo. In order to maintain the even distribution of the points, such a map $\Phi$ needs to be volume preserving in the sense that
\begin{align}
 \int_{\Omega \subseteq [0,1)^k} \mathrm{d}^k x = \frac{1}{\mathrm{Vol}(M)} \int_{\Phi(\Omega) \subseteq M} \mathrm{d}V_M
 \label{eq:fibAreaPres}
\end{align}
holds for all measurable sets $\Omega$.

To find such a map for $S_3$ (and therefore \SUTwo) we start by
introducing spherical coordinates
\begin{equation}
    z (\psi, \theta, \phi) = \matr{l}{
    \cos \psi\\
    \sin \psi \cos \theta\\
    \sin \psi \sin \theta \cos \phi\\
    \sin \psi \sin \theta \sin \phi}
\end{equation}
with
\[
 \psi \in [0,\pi)\,,\ \theta \in [0,\pi)\,,\ \phi \in [0,2\pi)\,.
\]
Therefore, the metric tensor $g_{ij}$ in terms of the spherical coordinates $(y_1, y_2, y_3) \coloneqq (\psi, \theta, \phi)$ is given by
\begin{align*}
 g_{ij} & = \frac{\partial z^a}{\partial y^i} \frac{\partial z^b}{\partial y^j} \delta_{ab} = \begin{pmatrix}
  1 & 0           & 0                         \\
  0 & \sin^2 \psi & 0                         \\
  0 & 0           & \sin^2 \psi \sin^2 \theta \\
 \end{pmatrix}_{ij} \textrm{.}
\end{align*}
From this one can calculate the Jacobian $\sqrt{\abs{g}}$ to be
\begin{align*}
 \sqrt{\abs{g}} & = \sin^2 \psi \sin \theta \textrm{.}
\end{align*}
As $\sqrt{\abs{g}}$ factorizes nicely into functions only dependent on one coordinate, one can construct a bijective map $\Phi^{-1}$ mapping $S_3$ to $[0,1)^3$ given by $\Phi^{-1}(\psi,\theta,\phi) = \left(\Phi_1^{-1} (\psi), \Phi_2^{-1}(\theta), \Phi_3^{-1} (\phi) \right)$ with
\begin{alignat*}{2}
 \Phi_1^{-1} (\psi)   & = \frac{\int_0^{\psi}\textrm{d}\tilde{\psi} \sin^2 \tilde{\psi}}{\int_0^\pi \textrm{d} \tilde{\psi} \sin^2 \tilde{\psi}} &  & = \frac{1}{\pi}  \left( \psi - \frac{1}{2} \sin( 2 \psi) \right) \\
 \Phi_2^{-1} (\theta) & = \frac{\int_0^{\theta}\textrm{d}\tilde{\theta} \sin \theta}{\int_0^\pi \textrm{d}\tilde{\theta} \sin \theta}            &  & = \frac{1}{2} \left( 1-\cos(\theta) \right)                      \\
 \Phi_3^{-1} (\phi)   & = \frac{\int_0^{\phi}\textrm{d}\tilde{\phi} }{\int_0^{2 \pi} \textrm{d} \tilde{\phi}}                                    &  & = \frac{1}{2 \pi} \phi \textrm{.}
\end{alignat*}
Looking at some measurable set $\Omega = \Phi^{-1} (\tilde{\Omega})$ one can
see that the inverse map $(\Phi^{-1})^{-1} \equiv \Phi$ trivially fulfils
equation \cref{eq:fibAreaPres}. A Fibonacci-like lattice on $S_3$
is therefore be given by
\begin{align*}
 F_n & = \left\{ z\left(\psi_m(t_m^1), \theta_m(t_m^2), \phi_m(t_m^3)\right)  \middle| \, 0 \le m < n, \, \, m \in \mathbb{N} \right\}\,,
\end{align*}
with
\[
\begin{split}
  \psi_m(t_m^1) & =  \Phi_1 \left( t_m^1 \right) = \Phi_1 \left(
  \frac{m}{n}\right)\,,\\
  \theta_m(t_m^2) & =  \Phi_2\left(t_m^2 \right) =
  \cos^{-1}\left(1-2(m\sqrt{2}\mod 1)\right)\,, \\
  \phi_m (t_m^3) & =  \Phi_3( t^3_m) = 2 \pi (m\sqrt{3} \mod 1)\,.
\end{split}
\]

\section{Methods}

\begin{figure}
  \includegraphics[width=\linewidth]{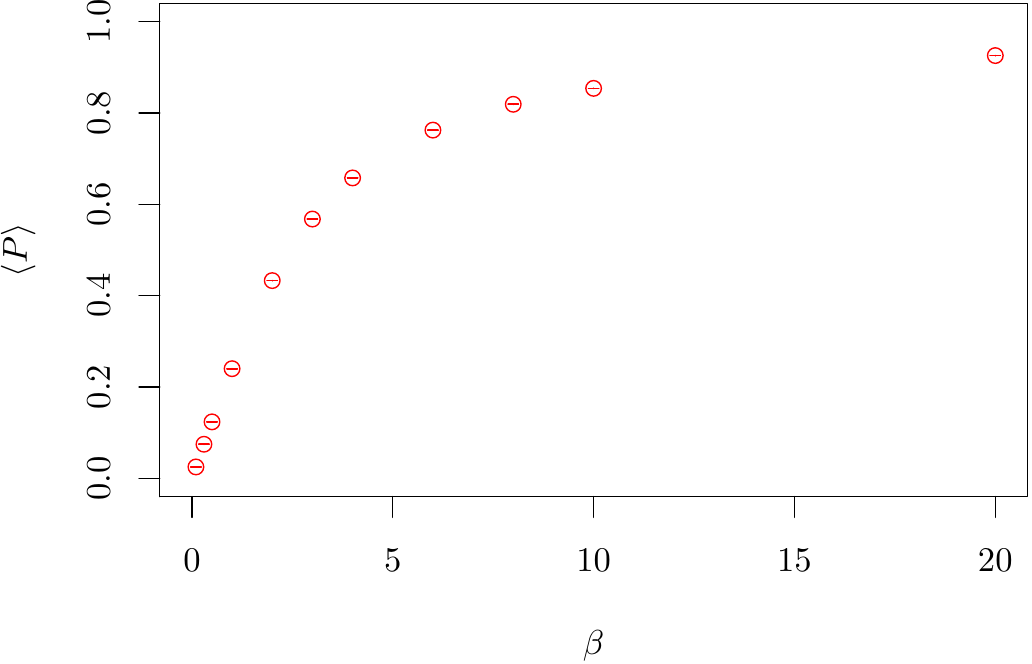}
  \includegraphics[width=\linewidth]{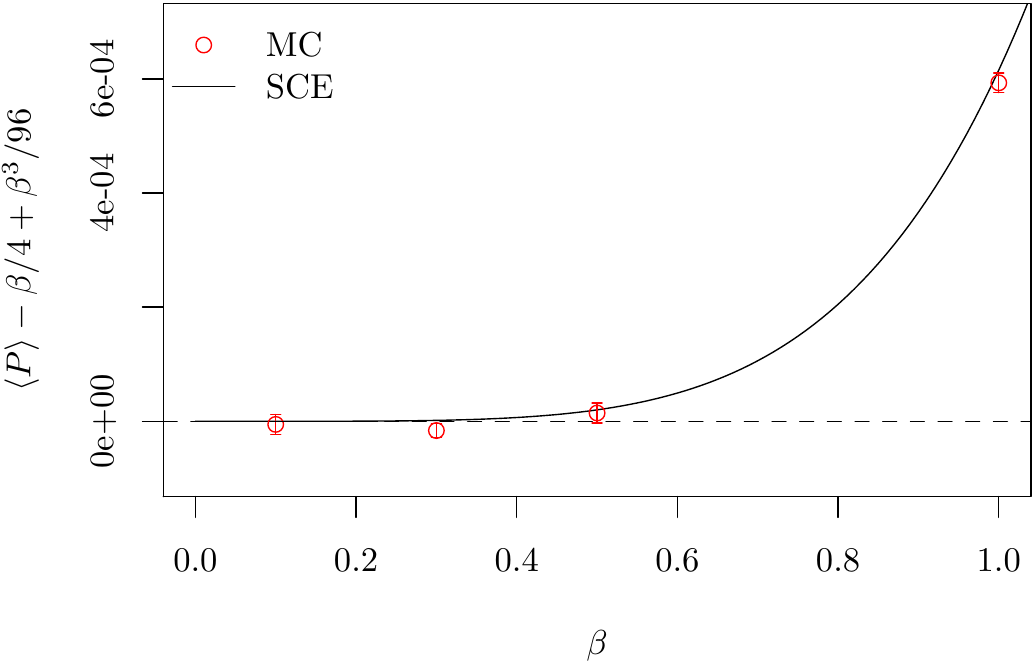}
  \caption{Plaquette expectation value as a function of $\beta$ in
    $1+1$ dimensions. In the lower panel we compare to the strong
    coupling expansion (SCE) \cref{eq:sce}.}
  \label{fig:PlaquetteVsBeta}
\end{figure}

In order to test the performance of the finite subgroups and the
partitionings discussed in the last section in Monte Carlo
simulations, we use a standard Metropolis Monte Carlo algorithm. It
consists of the following steps at site $n$ in direction $\mu$
\begin{enumerate}
\item generate a proposal $U_\mu'(n)$ from $U_\mu(n)$.
\item compute $\Delta S = S(U_\mu'(n)) - S(U_\mu(n))$.
\item accept with probability
  \begin{equation}
    \label{eq:Pacc}
    \mathbb{P}_\mathrm{acc} = \min\left\{1,\ \exp(-\Delta
    S)\frac{w(U_\mu'(n)}{w(U_\mu(n))}\right\} \,.
  \end{equation}
\end{enumerate}
This procedure is repeated $N_\mathrm{hit}$ times per
$n$ and $\mu$ before it is repeated for all $(n,\mu)$ pairs.
As reference we use an algorithm based on the double precision
floating point representation of the two complex elements $a,b$ needed
to represent an SU$(2)$ matrix
\begin{equation}
  U =
  \begin{pmatrix}
    a & b\\
    -b^\star & a^\star\\
  \end{pmatrix}
\end{equation}
with the additional constraint $aa^\star + bb^\star = 1$. In this case
$w(U) =1\,\forall\ U$ and the proposal is generated via $U_\mu'(n) =
V\cdot U_\mu(n)$, as explained above.
The algorithm can be tested for instance in the strong coupling limit
$\beta\to0$ against the strong coupling expansion derived in
Refs.~\cite{Balian:1974xw,PhysRevD.19.2514}, which reads in $d$
dimensions for the plaquette expectation value
\begin{equation}
  \label{eq:sce}
  \begin{split}
    \langle P\rangle&(\beta) = \frac{1}{4}\beta -
    \frac{1}{96}\beta^3 + \left(\frac{d}{96} -
    \frac{5}{288}\right)\frac{3}{16}\beta^5\\
    &+ \left(-\frac{d}{96} + \frac{29}{1440}\right)\frac{1}{16}\beta^7 +
    \mathcal{O}(\beta^9)\,.\\
  \end{split}
\end{equation}
In the upper panel of \cref{fig:PlaquetteVsBeta} we show the plaquette
expectation value as a function of $\beta$ in $d=1+1$ dimensions. In
the lower panel we compare to the corresponding strong coupling
expansion and find very good agreement.

For the subgroups the proposal step is implemented by multiplication
of $U_\mu(n)$ with one of the elements $V$ of the subgroup adjacent to
the identity. Also in this case the weights $w$ are constant.

In the case of the Genz points, the linear discretisation and the
Volleyball neighbouring points in the partitioning can be found by
geometric considerations as explained in the previous section. A
proposal is chosen uniformly random from the set of neighbouring
points. For the Genz points we do not take the weights $w$ into
account, because of their complex dependence on the direction and the
point itself. For the linear and the Volleyball discretisations we
compare simulations with and without taking the approximate weights
\cref{eq:linweights,eq:Vweights} into account.

Due to the locally irregular structure of the Fibonacci lattices
finding the appropriate neighboring elements is not as straightforward
as in the case of the other partitionings. Therefore, we pregenerate a
neighbour list for each element of the Fibonacci set based on the
geometric distance. This list is read and used during the update process.

In order to study the freezing transition, we follow the following
procedure. For a given $\beta$-value we perform a hot (random gauge
field) and a cold (unit gauge field) start separately. This is repeated
for $\beta$-values from $\beta_i\ll 1$ to $\beta_f$ in steps of
$\Delta\beta$. The phase transition is indicated by either the fact
that hot and cold starts do not equilibrate to the same average
plaquette expectation value for $\beta\geq\beta_c$ with one of the
two, typically the cold start, deviating from the reference result. Or
a significant deviation from the reference result is seen for
$\beta\geq\beta_c$ for both, hot and cold start.

For the purpose of this paper we define the critical value of $\beta$,
denoted as $\beta_c$ as the smallest value of $\beta$ for which forward
and backward branches do not agree within errors. In practice, this
will only be a lower bound for $\beta_c$. 

Statistical errors are computed based on the so-called $\Gamma$-method
detailed in Ref.~\cite{Wolff:2003sm} and implemented in the publicly
available software package hadron~\cite{hadron:2020}.

\section{Results}

\subsection{Influence of weights}

One important difference between finite subgroups and the
partitionings discussed above is the need for weights in the case of
the partitionings. In order to study the influence of weights, we
compare here Genz points with the linear discretisation for simplicity
in $d=1+1$ dimensions for $L^2=100^2$ lattices.

\begin{figure}
  \includegraphics[width=\linewidth]{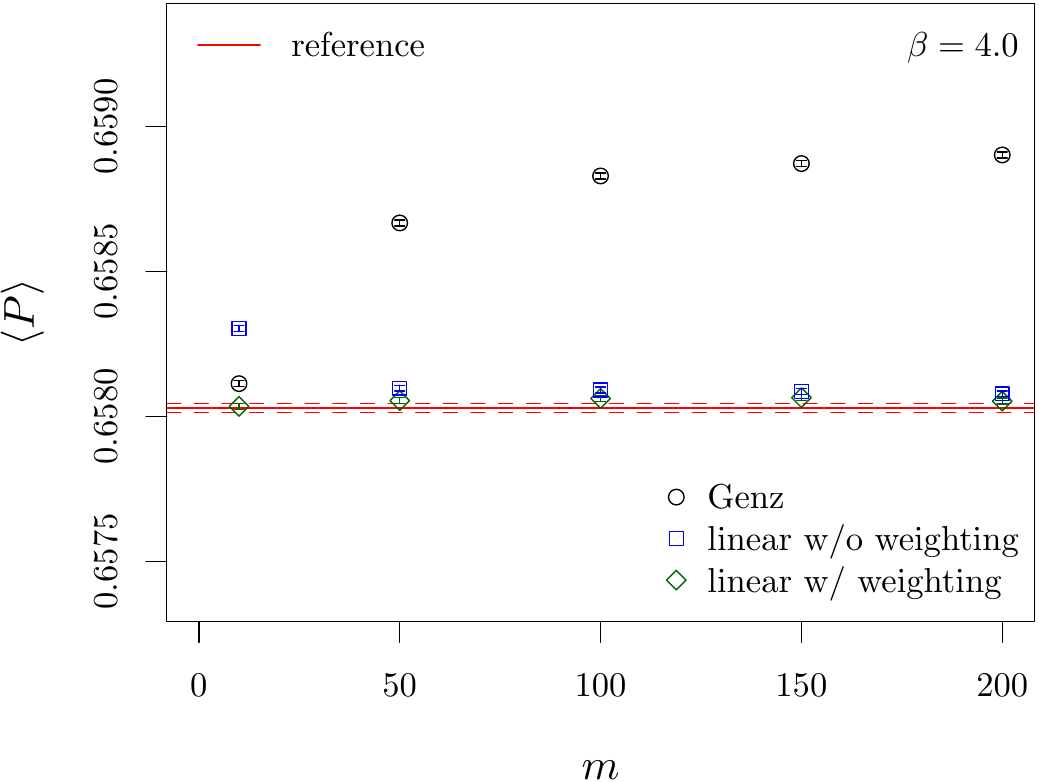}
  \caption{Comparison of plaquette expectation values for Genz
    partitioning $G_m$ and linear partitioning $L_m$ w/ and w/o
    weighting in $1+1$ dimensions on a $L^2=100^2$ lattice for
    $\beta=4.0$ as a  function of $m$.}
  \label{fig:genzvslin}
\end{figure}

In \cref{fig:genzvslin} we compare the plaquette expectation value
obtained from MC simulations with Genz points to those with the linear
discretisation with and without weighting taken into account for
$\beta=4.0$. The comparison is performed for values of $m$ in the
range from $5$ to $200$, which adjusts the fineness of the
partitioning. The reference result -- generated with the reference
algorithm as discussed above, is indicated by the red solid line
and the corresponding statistical uncertainty by the dashed red
lines. This $\beta$-value is representative. Only at very small
$\beta$-value, no dependence on $m$ can be observed.

We observe for the Genz points the strong
influence of missing weights. As expected from our estimate of the
weights, the deviation from the reference result increases with
increasing $m$.

In contrast, the linear discretisation without weighting converges
towards the reference result with increasing $m$. The smallish
deviations from the reference result at small $m$-values can be
reduced significantly (if not removed completely) by including the
weights in the MC simulation. This observation appears is
largely independent of $\beta$.

We conclude from these results that it is not worthwhile to further
consider the Genz points. For the linear partitioning it turns out
that the weights appear to be important for small $m$-values, but
become negligible for large $m$. However, this might also depend on
the observable.

\begin{figure}
  \centering
  \includegraphics[width=\linewidth]{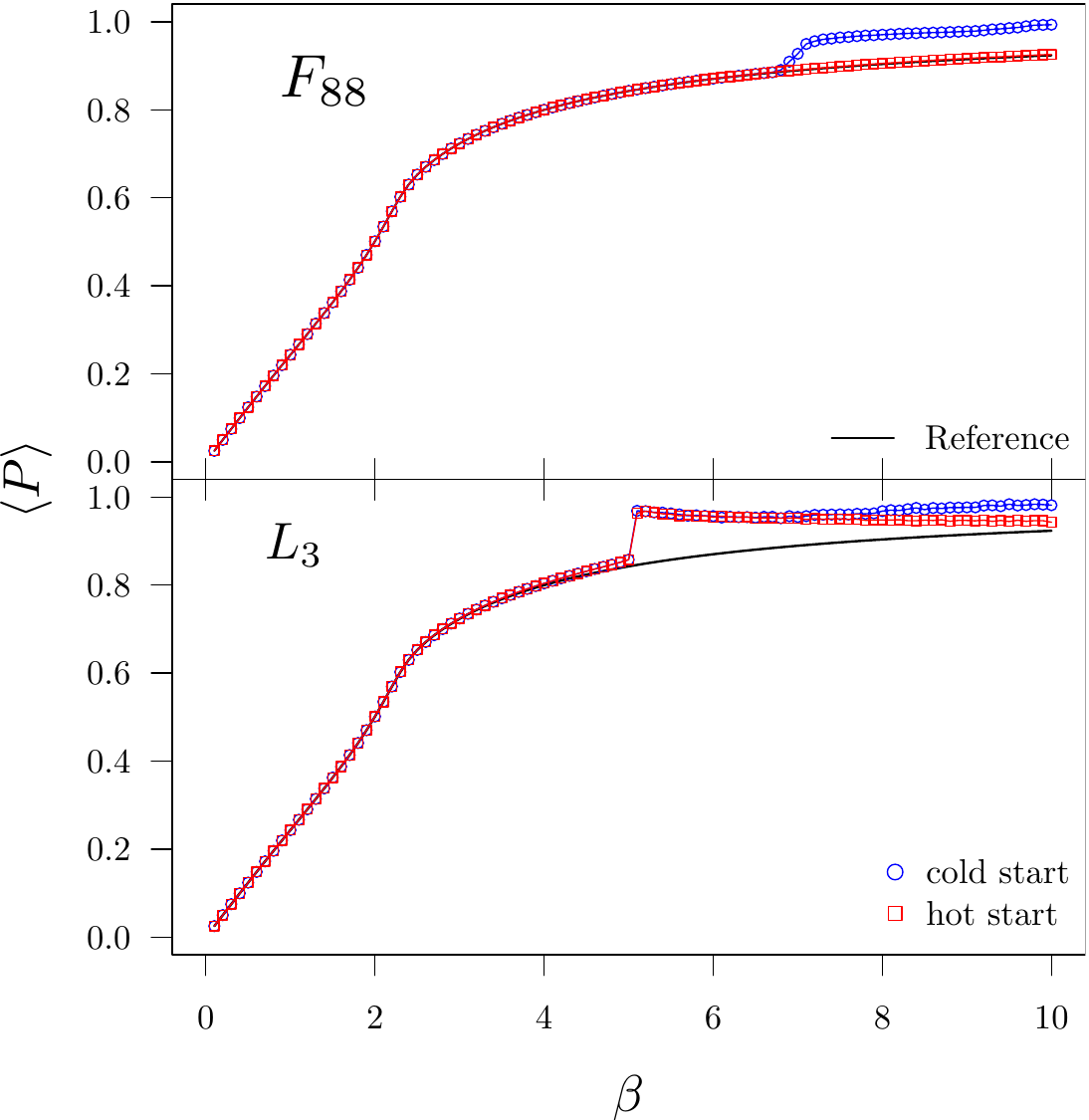}
  \caption{Hysteresis loops for the Fibonacci partitioning $F_{88}$
    and the linear partitioning with weights included $L_3$. Both have
  $n=88$ elements.}
 \label{fig:betaScanOverview}
\end{figure}

\subsection{Freezing Transition}

We study the freezing transition using simulations of the SU$(2)$
gauge theory in $3+1$ dimensions with
$L^4=8^4$ lattice volume. We look at $\beta \in \{ 0.1,0.2,\dots,
9.9,10.0\}$. For each value of $\beta$, $7000$ sweeps are performed,
once with a hot, and once with a cold starting configuration. During
a single sweep every lattice site and direction is probed
$N_{\mathrm{hit}} = 10$ times. The plaquette is then measured by
averaging over the last $3000$ iterations.

Such scans in $\beta$ can be found in \cref{fig:betaScanOverview}.
$\beta_c$ is then estimated to be the last value before a significant
jump in $\langle P \rangle$, or a significant disagreement between
the hot and cold start\footnote{Fibonacci lattices
usually show the latter behavior, which is why we increase the
number of thermalization sweeps to $\num{10000}$. This marginally
raised the values of $\beta_c$.}. We have checked that the
such determined critical $\beta$-values do not depend significantly
on the volume.

\begin{figure}
  \centering
  \includegraphics[width=\linewidth]{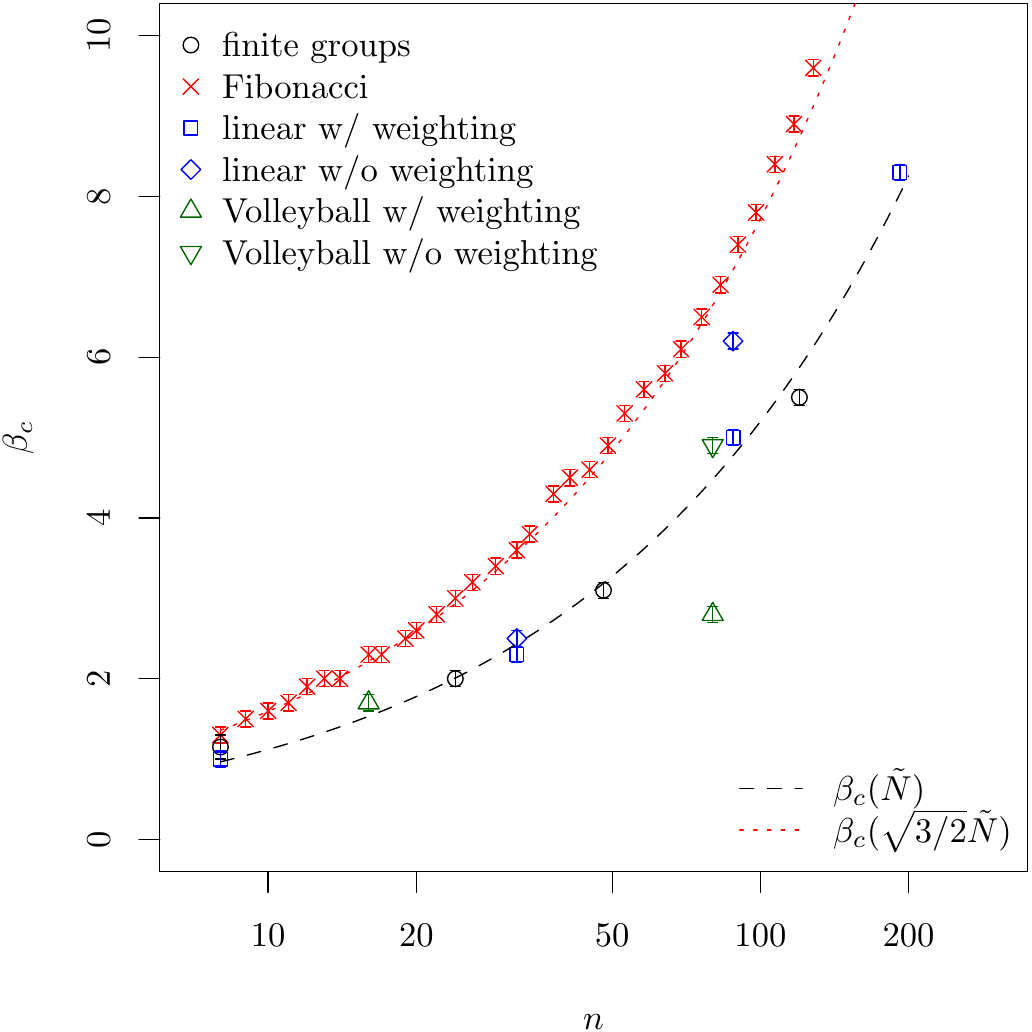}
  \caption{The critical value $\beta_c$ as a function of the number
    $n$ of elements in the set. The lines represent the
    approximation~\cref{eq:betac} where the order $\tilde{N}(n)$ is obtained
    from \cref{eq:approx_order}.}
 \label{fig:fibPhaseScan}
\end{figure}

In \cref{fig:fibPhaseScan} we show the $\beta_c$-values at which the
freezing transitions takes place as a function of the number $n$ of
elements in the set of points or the subgroups. We compare the
Fibonacci, the linear and the Volleyball partitioning, and
the finite subgroups of SU$(2)$. For the linear and the Volleyball
partitioning we also distinguish between results with and without
including weighting to correct for the different Voronoi cell
volumes. The corresponding results are also
tabularised in \cref{tab:betacfib,tab:betaclin,tab:betacSG}. For the
Fibonacci partitioning we tabularise only results for selected $n$
values.

Also note that our $\beta_c$-values for the finite subgroups
$\overline{T}$, $\overline{O}$ and $\overline{I}$ reproduce the ones
given in Ref.~\cite{Petcher:1980cq}.

\Cref{fig:fibPhaseScan} suggests that all our SU$(2)$ discretisations
behave qualitatively similar. However, at fixed $n$-value the
subgroups, linear and Volleyball partitionings do have smaller
$\beta_c$-values compared to the Fibonacci lattice. Moreover, we
observe a significant difference between simulations
with and without weighting included. This difference increases with
increasing $n$.

In Ref.~\cite{Petcher:1980cq} the authors find that the critical
$\beta$-value can be computed theoretically, at least approximately,
for the finite subgroups. It is based on an analytical calculation of
$\beta_c(N)$ for $Z_N$, which is generalised to finite subgroups as
follows: for the subgroup $G$, the authors define the set of
elements $C(G)$ closest to the identity, but excluding the identity
itself. Close to the freezing transition, plaquettes are made of
identity links, or $g, g^{-1}\in C(G)$ for minimal changes compared to
unit plaquettes. Next, they define $N$ (the cyclic order) as the
minimal integer for which
$g^N=1$. The corresponding subgroup generated by $g$ is isomorphic to
$Z_N$. For the four groups $\bar{D}_4$, $\overline{T}$, $\overline{O}$ and
$\overline{I}$ one finds $N=4,6,8$ and $10$, respectively.
This leads to the following expectation for the critical $\beta$-value
as a function of $N$
\begin{equation}
  \label{eq:betac}
  \beta_c(N) \approx \frac{\ln\left(1+\sqrt{2}\right)}{1-\cos(2\pi/N)}\,.
\end{equation}
However, for Fibonacci, linear and Volleyball partitionings we no
longer deal with subgroups. In particular, taking one of the elements
$e$ closest to the identity element, it is not guaranteed that there is an
$N\in\mathbb{N}$ for which $e^N=1$.

Thus, we have to approximate the order $N$. For (approximately)
isotropic discretisations such as the finite subgroups and the
Fibonacci partitioning a global average over the point density is
bound to yield a good approximation for the elements in $C(G)$ and
therefore $N$. The volume of the three dimensional unit sphere is
$2\pi^2$. If we then assume a locally primitive cubic lattice,
the average distance of $n$ points in $S_3$ becomes
\begin{equation}
  d(n) = \left(\frac{2\pi^2}{n}\right)^{1/3}\,.\label{eq:average_distance}
\end{equation}
Two points of this distance together with the origin form a triangle
with the opening angle
\begin{equation}
  \alpha(n) = 2\arcsin\frac{d(n)}{2}\,,
\end{equation}
thus a first approximation of the cyclic order is obtained by
\begin{equation}
  \label{eq:approx_order}
  \tilde{N}(n) = \frac{2\pi}{\alpha(n)}\,,
\end{equation}
which solely depends on the number $n$ of elements in the partition.

Note that the assumption of a primitive cubic lattice is even asymptotically incorrect for all the partitionings discussed in this work and at best a good approximation. How good an approximation it is, can only be checked numerically. In specific cases it needs further refinement. 

In particular, in the case of the Fibonacci partitioning the approximation has
to be adjusted. Since the points are distributed irregularly in this
case, a path going around the sphere does not lie in a two-dimensional
plane. Instead it follows some zigzag route which is longer than the
straight path. Assuming the optimal maximally dense packing, we expect
the points to lie at the vertices of tetrahedra locally tiling the
sphere. The length of the straight path would then correspond to the
height of the tetrahedron whereas the length of the actual path
corresponds to the edge length. Their ratio is $\sqrt{\frac32}$, so
$\tilde N$ has to be rescaled by this factor to best describe
$\beta_c$ for the Fibonacci partitioning.

We show the curve~\cref{eq:betac} using $\tilde{N}(n)$ and
$\sqrt{3/2}\tilde{N}$, respectively, in addition to the data in
\cref{fig:fibPhaseScan}. The version with $\tilde{N}$ is in very good
agreement with the results obtained for the finite subgroups while the
rescaled version matches the values for the Fibonacci partitioning
remarkably well.

The unweighted simulations of the Volleyball and the weighted
simulations of the linear discretisations also yield results
compatible with the unscaled version of \cref{eq:betac,eq:approx_order}.
On the other hand, the weighted Volleyball and the unweighted linear
discretisations deviate clearly.

\begin{table}[th]
  \centering
  \begin{tabular*}{.4\linewidth}{@{\extracolsep{\fill}}crr}
    \hline\hline
      $\Bigl. n$ & $\tilde{N}$ & $\beta_c$\\
    \hline
    8   & 4.2 & 1.4(1)\\
    12  & 5.0 & 1.9(1)\\
    24  & 6.4 & 3.1(1)\\
    32  & 7.1 & 3.7(1)\\
    53  & 8.5 & 5.1(1)\\
    88  & 10.2 & 6.8(1)\\
    90  & 10.3 & 7.0(1)\\
    152 & 12.3 & 9.6(1)\\
    \hline
  \end{tabular*}
  \caption{%
    $\beta_c$-values for selected Fibonacci lattice partitionings
    of SU$(2)$ for $d=4$ and $8^4$ lattices. Orders $\tilde N$ are
    approximations according to \cref{eq:approx_order},
    rounded to one digit.
  }
  \label{tab:betacfib}
\end{table}

\begin{table}[th]
  \centering
  \begin{tabular*}{.8\linewidth}{@{\extracolsep{\fill}}ccrrr}
    \hline\hline
      $\Bigl. m$ & $n$ & $\tilde{N}$ & $\beta_c^w$ & $\beta_c^{nw}$\\
    \hline
    1 & 8   & 4.2 & 0.9(1) & 0.9(1)\\
    2 & 32  & 7.1 & 2.3(1) & 2.5(1)\\
    3 & 88  & 10.2 & 5.0(1) & 6.2(1)\\
    4 & 192 & 13.3 & 8.3(1) & $>10$\\
    5 & 360 & 16.4 & $>15$ & \\
    \hline
  \end{tabular*}
  \caption{%
    $\beta_c$-values for the weighted and not-weighted linear
    discretisation $L_m$ for $1\leq m\leq  5$ determined for $d=4$ on $10^4$
    lattices.  Orders $\tilde{N}$ are approximations according to
    \cref{eq:approx_order}, rounded to one digit.
  }
  \label{tab:betaclin}
\end{table}

\begin{table}[th]
  \centering
  \begin{tabular*}{.8\linewidth}{@{\extracolsep{\fill}}lcccr}
    \hline\hline
    $\Bigl. $Subgroup & $n$ & $N$ & $\tilde{N}$ & $\beta_c$\\
    \hline
    $\bar{D}_4$ & 8 & $4$ & $4.2$ & 1.15(15)\\
    $\overline{T}$ & 24 & $6$ & $6.4$ & 2.15(15)\\
    $\overline{O}$ & 48 & $8$ & $8.2$ & 3.20(10)\\
    $\overline{I}$ & 120 & $10$ & $11.3$ & 5.70(20)\\
    \hline
  \end{tabular*}
  \caption{%
    $\beta_c$-values for the discrete subgroups of SU$(2)$. $N$ are the exact cyclic orders and $\tilde{N}$ are approximations according to \cref{eq:approx_order}.
  }
  \label{tab:betacSG}
\end{table}

\begin{table}[th]
  \centering
  \begin{tabular*}{.7\linewidth}{@{\extracolsep{\fill}}lccrr}
    \hline\hline
    $\Bigl. m$ & $n$ & $\tilde{N}$ & $\beta_c^w$ & $\beta_c^{nw}$\\
    \hline
    0 & 16 & 5.6 & 1.7(1) & 1.7(1)\\
    1 & 80 & 9.8 & 2.8(1) & 4.9(1)\\
    \hline
  \end{tabular*}
  \caption{%
    $\beta_c$-values for the weighted and not-weighted Volleyball
    discretisation $V_m$ for $0\leq m\leq  1$ determined for $d=4$ on $10^4$
    lattices.  Orders $\tilde N$ are approximations according to \cref{eq:approx_order}, rounded to one digit.
  }
  \label{tab:betacVol}
\end{table}

\section{Discussion and Outlook}

Some of the results presented in the previous section deserve separate
discussion. \Cref{fig:fibPhaseScan} shows that the Fibonacci lattice
discretisation has larger $\beta_c$-values at fixed $n$ compared to
the finite subgroups and the other partitionings. This can be
understood due to the irregularity of the points in the Fibonacci
lattices: at fixed $n$, this irregularity will generate minimal
distances between points which are smaller than the ones for the other
discretisations. Thus, the freezing transition should appear at
comparably larger $\beta$ values only, because smaller values of
$|\Delta S|$ are available.

Also the difference in $\beta_c$ between simulations with and without
weight included for the linear and Volleyball partitionings,
respectively, can be understood qualitatively. Assume the system
freezes for the weighted case at some value $\beta_c^w$.
Switching off the weighting, there will be subsets of points
with on average lower (or larger) distances between elements than the
average distance. In these regions the $|\Delta S|$ values required
for acceptance will be smaller than the average $|\Delta S|$ value at
this $\beta$. And it is reasonable to assume that these regions are
also reached during equilibration. Thus, the critical $\beta$-value
for the not weighted simulation $\beta^{nw}$ must be larger or equal
$\beta^w$.

Though this trend is universal, we find an additional superiority of
the linear as compared to the Volleyball discretisation. We expect
this to be a consequence of the denser packing of the linear
discretisation where most of the points have twelve neighbours
whereas the majority of the points in the  
Volleyball discretisation has only six neighbours.

We have obtained excellent predictions for the
$\beta_c$-values for finite subgroups and Fibonacci partitionings. For
finite subgroups the prediction using $\tilde{N}$ is even better than
the prediction using $N$, in particular for larger $n$. For the
Fibonacci partitionings the rescaling with the factor $\sqrt{3/2}$
suggests that the Fibonacci elements are close to maximally densely
packed. This is strongly backed up by the numerical evidence.
Based on this assumption of closest-packing we 
postulate that there is no discretisation scheme yielding a significantly later 
freezing transition than the Fibonacci partitioning at an equal number of 
points.

Finally, the predicted $\beta_c$ values do not agree with our
observations for the unweighted linear and the weighted Volleyball
discretisations, respectively. We do not fully understand these
discrepancies, but it suggests that the linear discretisation has
subsets of elements which are close to maximally densely packed. And
the Volleyball discretisation is in this regard sub-optimal.

\begin{figure}
	\centering
	\includegraphics[width=\linewidth]{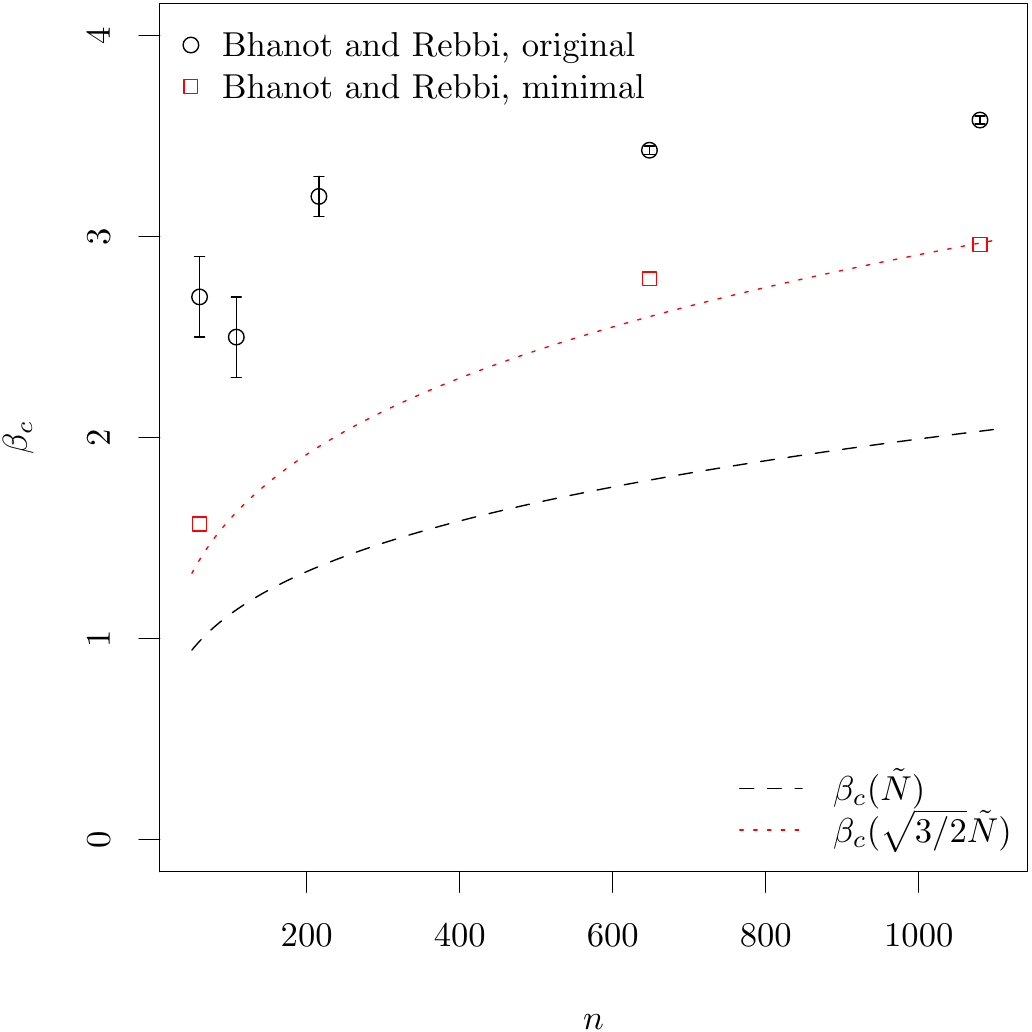}
	\caption{The critical value $\beta_c$ in SU$(3)$ as a function of the number
		$n$ of elements in the set. The lines represent the
		approximation~\cref{eq:betac} where the order $\tilde{N}(n)$ is obtained
		from \cref{eq:average_SUN_distance} and \cref{eq:approx_order}. The reference data by Bhanot and Rebbi~\cite{Freezing_SU3} comes from their Table~1 (``original'') and from the leftmost points of the hysteresis loops in their Figures~1-3 (``minimal''), respectively.}
	\label{fig:su3PhaseScan}
\end{figure}

In \Cref{sec:dense_partitionings} we have explained how SU$(N)$ can be expressed as a product of odd-dimensional spheres $S_3$, $S_5$, $\ldots$, $S_{2N-1}$. Since the $k$-dimensional hypervolume $H(S_k)\equiv2\pi^{\frac{k+1}{2}}/\Gamma\left(\frac{k+1}{2}\right)$ of the $k$-sphere is well known, we can generalise the prediction of $\beta_c$ to $N>2$ by adjusting the average distance from \cref{eq:average_distance} 
\begin{align}
	d_N(n) = \left(\frac1n\prod_{k=3,\text{odd}}^{2N-1}H(S_k)\right)^{1/(N^2-1)}\label{eq:average_SUN_distance}
\end{align}
and applying \cref{eq:approx_order} and \cref{eq:betac} successively as before.
In particular, this formula readily predicts critical couplings for the finite subgroups of SU$(3)$ which have been determined by Bhanot and Rebbi~\cite{Freezing_SU3}. We show how our prediction compares to the values obtained by Bhanot and Rebbi in \cref{fig:su3PhaseScan}. In addition to the results for $\beta_c$ given in the paper originally (black circles), we plot the leftmost bounds of the hysteresis loops\footnote{The hysteresis loop is formed by the different values of the plaquette corresponding to hot and cold starts, respectively. The minimal $\beta_c$ is the first point where these results do not agree.} they visualised, denoting the minimal possible value of $\beta_c$ (red squares). The systematic effect stemming from different estimations of $\beta_c$ is remarkably large. We therefore refrain from any conclusion as to the quantitative correctness of our prediction. Nevertheless it seems well suited to predict the qualitative scaling of the freezing transition and it provides the correct order of magnitude.

\section{Summary}

In this paper we have presented several asymptotically dense
partitionings of SU$(2)$, which do not represent subgroups of SU$(2)$
but which have adjustable numbers of elements. The discussed
partitionings are not necessarily isotropically distributed in the
group, which requires in principle the inclusion of additional weight
factors in the Monte Carlo algorithms. We have investigated
whether or not the partitionings without and, if possible, with
weights included can be used in Monte Carlo simulations of SU$(2)$
lattice gauge theories by comparing the plaquette expectation value as
a function of $\beta$ to reference results of a standard lattice gauge
simulation.

This comparison rules out the usage of the so-called Genz
partitioning, because the weights are difficult to compute and the
difference due to not included weights increases with the number of
elements. Thus, Monte Carlo simulations with fine Genz partitionings
of SU$(2)$ are not feasible.

For the other considered partitionings, this comparison turned out to give
good agreement with the standard simulation code, in particular when
the weights are included. Moreover, the finer the discretisation (and
the larger the number of elements) the smaller the deviation between
simulations with and without weighting.

In addition we have investigated the so-called freezing transition for
the partitions and for all finite subgroups of SU$(2)$. The main
result visualised in \cref{fig:fibPhaseScan} is that the partitioning
$F_k$ based on Fibonacci lattices allows for a flexible choice of the
number of elements by adjusting $k$ and at the same time larger
$\beta_c$-values compared to finite subgroups and the other discussed
partitionings. Thus, Fibonacci based discretisations provide the
largest simulatable $\beta$-range at fixed $n$.

Coming back to the introduction, using the partitionings proposed here
does not pose any problem even at very large $\beta$-values at least
in Monte Carlo simulations. This leaves us optimistic for their
applicability in the Hamiltonian formalism for tensor network or
quantum computing applications.

Finally, the generalisation of the partitionings discussed here to
the case of SU$(3)$ relevant for quantum chromodynamics is
straightforward and we expect that the results obtained in this paper
directly translate to this larger group.

\begin{acknowledgments}
  This work is supported by the Deutsche
  Forschungsgemeinschaft (DFG, German Research Foundation) and the  
  NSFC through the funds provided to the Sino-German
  Collaborative Research Center CRC 110 “Symmetries
  and the Emergence of Structure in QCD” (DFG Project-ID 196253076 -
  TRR 110, NSFC Grant No.~12070131001).
  The open source software packages R~\cite{R:2019} and
  hadron~\cite{hadron:2020} have been used.
\end{acknowledgments}

\bibliography{bibliography}

\onecolumngrid
\appendix

\section{Derivations of the nearest neighbour distances}\label{sec:distances_derivation}

The square distance $d(j_i, j_l)^2$ for \textit{Genz points} can be
approximated as follows:
\begin{align}
  d(j_i,j_l)^2 &= \left|\left(\sqrt{\frac{j_i}{m}}-\sqrt{\frac{j_i-1}{m}},\,\sqrt{\frac{j_l}{m}}-\sqrt{\frac{j_l+1}{m}}\right)\right|^2\\
  &= \frac 1m \left(j_i+j_i-1-2\sqrt{j_i(j_i-1)}+j_l+j_l+1-2\sqrt{j_l(j_l+1)}\right)\\
  &= \frac 1m \left(2j_i-2\left(j_i-\frac12-\frac{1}{8j_i}+\ordnung{j_i^{-2}}\right)+2j_l-2\left(j_l+\frac12-\frac{1}{8j_l}+\ordnung{j_l^{-2}}\right)\right)\\
  &= \frac 1m \left(\frac{1}{4j_i}+\frac{1}{4j_l}+\ordnung{j_i^{-2}}+\ordnung{j_l^{-2}}\right)
\end{align}

For the \textit{linear partitioning} one readily computes the approximation
\begin{align}
  \Delta M &\equiv \sqrt{\sum_{i'=0}^kj_{i'}^2+(j_i-1)^2-j_i^2+(j_l+1)^2-j_l^2}-\sqrt{\sum_{i'=0}^kj_{i'}^2}\\
  &= \sqrt{M^2-2j_i+2j_l+2}-M\\
  &= \frac{-j_i+j_l}{M} + \ordnung{\frac 1M}+ \ordnung{\frac{\left(j_i-j_l\right)^2}{M^3}}\\
  &= \frac{j_l-j_i}{M} + \ordnung{\frac 1m}
\end{align}
yielding an inverse change
\begin{align}
  \frac 1M - \frac{1}{M+\Delta M} &= \frac{\Delta M}{M^2}+\ordnung{\frac 1{M^3}}\\
  &= \frac{j_l-j_i}{M^3} + \ordnung{\frac 1{m^3}}\,.
\end{align}
With this we can again calculate the square distance
\begin{align}
  \begin{split}
    d(j_i,j_l)^2 &= \left|\left(\frac{j_0}{M}-\frac{j_0}{M+\Delta M},\right.\right.\\
    &\qquad\dots,\frac{j_i}{M}-\frac{j_i-1}{M+\Delta M},\dots,\frac{j_l}{M}-\frac{j_l+1}{M+\Delta M},\\
    &\qquad\dots,\left.\left.\frac{j_k}{M}-\frac{j_k}{M+\Delta M}\right)\right|^2
  \end{split}\\
  &= \sum_{l=0}^k\left(\frac{j_l(j_l-j_i)}{M^3}\right)^2+\frac{2}{M^2}+\ordnung{\frac{1}{m^3}}\\
  &=\frac{(j_l-j_i)^2}{M^4}+\frac{2}{M^2}+\ordnung{\frac{1}{m^3}}\,.
\end{align}

Nearest neighbours for the \textit{Volleyball partitioning} can be
obtained by modifying $j_i \pm 1$ leading to the following
approximation of $\Delta M$
\begin{align}
  \Delta M &\equiv \sqrt{\sum_{i'=0}^kj_{i'}^2+(j_i \pm 1)^2-j_i^2} - \sqrt{\sum_{i'=0}^kj_{i'}^2}\\
  &= \sqrt{M^2 \pm 2j_i + 1} - M\\
  &= \frac{\pm j_i}{M} + \ordnung{\frac{1}{M}}+ \ordnung{\frac{j_i^2}{M^3}}\\
  &= \frac{\pm j_i}{M} + \ordnung{\frac 1m}
\end{align}
and the corresponding inverse change
\begin{align}
  \frac 1M - \frac{1}{M+\Delta M} &= \frac{\Delta M}{M^2}+\ordnung{\frac 1{M^3}}\\
  &= \frac{\pm j_i}{M^3} + \ordnung{\frac 1{m^3}}\,.
\end{align}
Finally, we find the square distance
\begin{align}
  \begin{split}
    d(j_i,&j_l)^2 = \left|\left(\frac{j_0}{M}-\frac{j_0}{M+\Delta M},\right.\right.\\
    &\dots,\frac{j_i}{M}-\frac{j_i\pm 1}{M+\Delta M},\dots,\left.\left.\frac{j_k}{M}-\frac{j_k}{M+\Delta M}\right)\right|^2
  \end{split}\\
  &= \sum_{l=0}^k\left(\frac{j_l j_i}{M^3}\right)^2+\frac{1}{M^2}+\ordnung{\frac{1}{m^3}}\\
  &=\frac{j_i^2}{M^4}+\frac{1}{M^2}+\ordnung{\frac{1}{m^3}}\,.
\end{align}

\end{document}